\documentclass[11pt]{article}
\usepackage[a4paper]{geometry}
\usepackage{amsmath}
\usepackage{amsthm}
\usepackage{amssymb}
\usepackage{graphicx}
\usepackage[vcentermath]{youngtab}
\usepackage{hyperref}
\usepackage{multicol,color,longtable}
\definecolor{darkred}{rgb}{0.65,0.15,0}
\definecolor{darkgreen}{rgb}{0.42, 0.56, 0.14}
\definecolor{darkblue}{rgb}{0.02, 0.20, 0.30}
\definecolor{darkcerulean}{rgb}{0.03, 0.27, 0.49}
\definecolor{oucrimsonred}{rgb}{0.65, 0.0, 0.0}
\hypersetup{pdfborder={0 0 0},colorlinks=true,urlcolor=darkblue,citecolor=darkcerulean,linkcolor=oucrimsonred,linktocpage=true}
\usepackage{float}
\restylefloat{table}

\def\ft#1#2{\tfrac{#1}{#2}}

\def\ie{{\it i.e.}\ }

\newcommand{\be}{\begin{equation}}
\newcommand{\ee}{\end{equation}}
\newcommand{\bea}{\begin{eqnarray}}
\newcommand{\eea}{\end{eqnarray}}
\newcommand{\ba}{\begin{array}}
\newcommand{\ea}{\end{array}}

\newfont{\bbbold}{msbm10}

\def\bbR{\mbox{\bbbold R}}

\begin{document}

\newcommand{\hoch}[1]{$\, ^{#1}$}
\newcommand{\imperial}{\it\small Theoretical Physics Group, Imperial College London\\ Prince Consort Road, London SW7 2AZ, UK}
\newcommand{\cpht}
{\it\small Centre de Physique Th\'eorique, CNRS,  Institut Polytechnique de Paris\\
91128 Palaiseau cedex, France}

\newcommand{\auth}{\large G.\ Bossard\footnote{email: guillaume.bossard@polytechnique.edu} and K.S.\ Stelle\footnote{email: k.stelle@imperial.ac.uk}}

\thispagestyle{empty}

\renewcommand{\thefootnote}{\arabic{footnote}}

\null
\begin{flushright}
{\small CPHT-RR020.022025\\
Imperial/TP/2025/KS/{01}}

\vskip 1 cm
\end{flushright}

\begin{center}
{\Large{\bf  The Ultraviolet Problem in Supergravity}}
\vspace{.75cm}

\auth

\vspace{.5cm}

\begin{itemize}
\item [$^1$]\cpht \item [$^2$] \imperial
\end{itemize}
\vspace{.4cm}

{\bf Abstract}
\end{center}
We review the development of understanding for the problem of ultraviolet divergences in supergravity. This history proceeds from initial constructions of counterterms invariant under the relevant degrees of local supersymmetry, through a deeper understanding of non-renormalisation theorems for the relevant degrees of ``off-shell'' linearly realisable supersymmetry, and on to current understanding of limitations on counterterm eligibility based on the duality symmetries of supergravity theories, as well as of the related structures emerging in superstring theory.


\vspace{5cm}

\noindent{\it Invited contribution to the volume ``Half a Century of Supergravity,'' eds. A.\ Ceresole\\ and G.\ Dall'Agata.}
\renewcommand{\thefootnote}{\arabic{footnote}}
\setcounter{footnote}{0}

\pagebreak

\section{UV divergences in gravity and supergravity}\label{intro}

A basic difficulty in formulating a quantum theory of gravity was already recognized in the earliest approaches in the 1930's: the dimensional character of Newton's constant gives rise to ultraviolet divergent quantum correction integrals. Na\"\i ve power counting of the degree of divergence $\Delta$ of an $L$-loop diagram in $D$-dimensional gravity theories yields the result
\be
\Delta = (D-2)L + 2\  ,
\label{degdiv}
\ee
which grows linearly with loop order, implying a requirement for higher and higher dimensional counterterms needed to renormalize the divergences.  In the 1970's, this was confirmed explicitly in the first Feynman diagram calculations of the radiative corrections to systems containing gravity plus matter \cite{'tHooft:1974bx}. The time lag between the general perception of the UV divergence problem and its first concrete demonstration was due to the complexity of Feynman diagram calculations involving gravity. The necessary techniques were an outgrowth of the long struggle to control, consistently with Lorentz invariance, the quantization of non-abelian Yang-Mills theories in the Standard Model of weak and electromagnetic interactions and likewise in quantum chromodynamics.

With the advent of supergravity \cite{Freedman:1976xh,Deser:1976eh} in the mid 1970's, hopes rose that the specific combinations of quantum fields present in supergravity theories might possibly tame the gravitational UV divergence problem. Indeed, it turns out that all the irreducible supergravity theories in four-dimensional spacetime, {\it i.e.}\ theories in which all fields are irreducibly linked to gravity by supersymmetry transformations, have remarkable cancellations in Feynman diagrams at low loop orders.

There is a sequence of such irreducible (or ``pure'') supergravity models, characterized by the number $N$ of local, that is, spacetime-dependent, spinor transformation parameters. In four-dimensional spacetime, minimal, {\it i.e.}\  $N=1$, supergravity thus has 4 supersymmetries corresponding to the components of a single Majorana spinor transformation parameter. The maximal possible supergravity \cite{Cremmer:1979up} in four dimensional spacetime has $N=8$ spinor parameters, {\it i.e.}\ 32 independent supersymmetries.

The hopes for ``miraculous'' UV divergence cancellations in supergravity were subsequently dampened, however, by the realization that the divergence-killing powers of supersymmetry most likely do not extend beyond the two-loop order for generic pure supergravity theories \cite{Deser:1977nt,Kallosh:1980fi,Howe:1981xy,Howe:1983sr}. The anticipated three-loop  invariant counterterm \cite{Deser:1977nt} is quartic in curvatures, and has a purely gravitational part given by the square of the Bel-Robinson tensor \cite{Deser:1977nt}:
\be
\int\sqrt{-g}T_{\mu\nu\alpha\beta}T^{\mu\nu\alpha\beta}\ ,\qquad
T_{\mu\nu\alpha\beta} = R^{\lambda\, \, \, \rho}_{\, \, \, \alpha \, \, \, \mu} \, R_{\lambda\beta\rho\nu} + \ ^*R^{\lambda\, \, \, \rho}_{\, \, \, \alpha \, \, \, \mu} \, ^*R_{\lambda\beta\rho\nu}\,.
\ee

The development of candidate locally supersymmetric counterterms became significantly advanced with the construction of linearly supersymmetric realisations for supergravity, beginning with the auxiliary-field formulation of minimal $N=1$, $D=4$ supergravity \cite{Stelle:1978ye,Ferrara:1978em}. This became further formalised with the development of superspace formulations of supergravity, which were intensely discussed at the 1981 Nuffield Supergravity Workshop in Cambridge \cite{Hawking:1981bu}. For extended-supersymmetry supergravities, however, the realisation of such ``off-shell'' superspace formulations becomes progressively more problematic, as demonstrated by ``no-go'' arguments, at least for supersymmetry formulations incorporating finite numbers of component fields \cite{Taylor:1981gr}. This applied particularly to the possibility of off-shell realisations of the full supersymmetry, but the picture is complicated by the existence of ``harmonic superspace'' formulations incorporating infinite numbers of component fields \cite{Galperin:2001seg}. 

Further study revealed the existence, of formulations of maximal $N=4$ super Yang-Mills and maximal $N=8$ with half-maximal superspace formulations (\ie $N=2$ for $D=4$ maximal super Yang-Mills \cite{Howe:1982tm,Mandelstam:1982cb,Brink:1982wv} and $N=4$ for maximal supergravity \cite{Howe:1982mt}). When coupled with non-renormalisation theorems \cite{Grisaru:1982zh,Howe:1983sr}, based on the background-field method of quantisation, that restrict superspace divergences to {\em full-superspace} integrals for the achievable off-shell supersymmetry, this gave an initial indication of the loop orders of Feynman diagrams at which the initial ultraviolet divergences could appear. 

Similarly to the way in which chiral integrals of $N=1$, $D=4$ supersymmetry achieve invariance in integrals over less than the theory's full superspace, provided that the integrand satisfies a corresponding BPS-type constraint, there are analogous invariants involving integration over varying portions of an extended supersymmetric theory's full superspace \cite{Howe:1981xy}. Thus, ``half-BPS'' operators require integration over just half the full set of fermionic $\theta$ coordinates. And if half the full supersymmetry were the maximal amount that is linearly realizable (with corresponding constraints on diagrams from the corresponding Ward identities), such operators would be the first to be allowed as UV counterterms \cite{Howe:1988qz}. The detailed analysis of which counterterms are allowed and which are not involves the extent to which counterterm integrands must be manifestly gauge invariant and must also respect the other rigid automorphism symmetries of a theory. For this purpose, the background field method \cite{Howe:1981xy} is an essential tool. It can be used to calculate effective action contributions with only background fields on external lines, and yet one can use Ward identities for the background-quantum split \cite{Howe:1986vm} to show how to renormalize all diagrams occurring at higher orders. In the case of supersymmetric gauge theories, this implies that counterterm integrands must be written in terms of background gauge connections, and must not involve prepotentials explicitly. This analysis of counterterm eligibility was backed up by agreement with the results of explicit Feynman diagram computations in maximal super Yang-Mills theory \cite{Marcus:1984ei}.

A main point of contention in the early analyses concerned the eligibility of the half-BPS supergravity counterterms, either at $L=2, D=5$ or at $L=3, D=4$. These have power-counting weight $\Delta=8$, and have generic structure $(\hbox{curvature})^4$ (which one may telegraphically denote $R^4$, directly continuing the structure of the first such ``$R^4$'' counterterm found for $N=1$ supergravity in \cite{Deser:1977nt}). The detailed structure at the leading quartic order in fields of this maximal supergravity counterterm \cite{Howe:1981xy} reveals its similarity to the analogous $(\hbox{field strength})^4$ candidate super Yang-Mills counterterm. Written in terms of $D=4$ on-shell linearized superfields, the candidate counterterms are written in terms of the basic scalar superfields carrying totally antisymmetric R-symmetry indices $\phi_{ij}$ carrying a {\bf 6} of $SU(4)$ in maximal super Yang-Mills or $W_{ijkl}$ carrying a {\bf 70} of $SU(8)$ in maximal supergravity:
\begin{eqnarray}
\Delta I_{\rm SYM} &=&\int d^4x (d^4\theta d^4\bar\theta)_{\bf 105}\,{\mathrm tr}(\phi^4)_{\bf 105}\qquad\qquad\quad\ \ \, {\bf 105}\leftrightarrow  {\Yboxdim7pt \yng(4,4)}\label{symf4}\\
\Delta I_{\rm SG}&=&\int d^4x (d^8\theta d^8\bar\theta)_{\bf 232848}(W^4)_{\bf 232848}\qquad {\bf 232848}\leftrightarrow  {\Yboxdim7pt \yng(4,4,4,4)}\label{sgr4}
\end{eqnarray}
These structures reveal explicitly their half-BPS character, involving integrations over just 8 of the 16 odd superspace coordinates for maximal super Yang-Mills and 16 of the 32 odd coordinates for maximal supergravity. 

These early studies consequently indicated that non-renormalization theorems deriving from superspace quantization might allow the first UV divergences at the loop orders shown in Table \ref{tab1} for various spacetime dimensions:

\begin{table}[htp]
\centering
\begin{tabular}{|l|c|c|c|c|c|c|c|}
\hline
Dimension $D$&11&10&8&7&6&5&4\\
\hline
Loop order $L$&2&2&1&2&3&2&3\\
\hline
General form&$\partial^{12} R^4$&$\partial^{10} R^4$&$R^4$&$\partial^4 R^4$&$\partial^6 R^4$&$R^4$&$R^4$\\
\hline
\end{tabular}
\caption{Early expectations for maximal supergravity first divergences, assuming half the supercharges ({\it i.e.}\ 16) are linearly realizable.
\label{tab1}}
\end{table}

These early divergence expectations involved combining a number of different requirements. These included the fraction of a given theory's supersymmetry that could be linearly (or "off-shell") realized in the Feynman rules, respect of gauge invariances as well as the availability of an invariant with respect to the full on-shell supersymmetry that remains nonvanishing subject to the classical equations of motion. (Only such divergences require counterterms; divergence structures that vanish subject to the classical field equations can be removed by field redefinitions.)

Despite the above UV outlook, a faint hope persisted among some researchers that the maximal $N=8$ supergravity might have very special UV properties, in distinction to the non-maximal cases. This hope was bolstered by the complete ultraviolet finiteness in $D=4$ dimensions of maximal $N=4$ supersymmetric Yang-Mills theory \cite{Howe:1982tm,Mandelstam:1982cb,Brink:1982wv}. This was the first interacting UV-finite theory in four spacetime dimensions. In higher dimensional spacetimes, super Yang-Mills theory itself becomes divergent, however, with the expected first divergence orders providing a very useful comparison to the supergravity cases. Based upon similar analysis to the supergravity cases, early expectations were that the initial divergences in super Yang-Mills theory could occur for various spacetime dimensions as shown in Table \ref{tab2}:

\begin{table}[htp]
\centering
\begin{tabular}{|l|c|c|c|c|c|c|}
\hline
Dimension $D$&10&8&7&6&5&4\\
\hline
Loop order $L$&1&1&2&3&4&$\infty$\\
\hline
General form&$\partial^2 F^4$&$F^4$&$\partial^2 F^4$&$\partial^2 F^4$&$F^4$&finite\\
\hline
\end{tabular}
\caption{Early expectations for maximal super Yang-Mills first divergences, assuming half the supercharges ({\it i.e.}\ 8) are linearly realizable.
\label{tab2}}
\end{table}

Of course, one cannot prove the presence of an ultraviolet divergence simply by studying nonrenormalization theorems. This requires a proper calculation. There is a ``folk expectation'', however, that in complicated Feynman diagram calculations, vanishing results do not happen without a clear underlying reason. But this does not strictly rule out the possibility of ``miraculous'' cancellations which are not predicted by the nonrenormalization theorems. But the prevailing expectation runs against the occurrence of such ``miracles''.

It is just such apparently ``miraculous'' UV divergence cancellations that became confirmed, however, in remarkable 3-loop and 4-loop calculations in maximal super Yang-Mills \cite{Bern:2006ew} and maximal supergravity \cite{Bern:2007hh,Bern:2009kd}.  Performing such calculations at high loop orders requires a departure from textbook Feynman-diagram methods \cite{Bern:1994zx}, because the standard approaches can produce astronomical numbers of terms. Instead of following the standard propagator \& vertex methods for supergravity calculations, Bern et al.\ used another technique which dates back to Feynman: loop calculations can be performed using the unitarity properties of the quantum S-matrix. These involve cutting rules that reduce higher-loop diagrams to sums of products of leading-order ``tree'' diagrams without internal loops. This use of unitarity is an outgrowth of the optical theorem in quantum mechanics for the imaginary part of the S-matrix.

In order to obtain information about the real part of the S-matrix, an additional necessary element in the unitarity-based technique is the extended use of dimensional regularization to render UV divergent diagrams formally finite. In dimensional regularization, the dimensionality of spacetime is changed from $4$ to $4-\epsilon$, where $\epsilon$ is a small adjustable parameter. Traditional Feynman diagram calculations also often use dimensional regularization, but normally one just focuses on the leading $1/\epsilon$ poles in order to subtract them in a renormalization program. In the unitarity-based approach, all orders in $\epsilon$ need to be retained. This gives rise to logarithms in which real and imaginary contributions are related.

In maximal $N=8$ supergravity theory, the complexity of the quantum amplitudes factorizes, with structures involving the particular field types occurring on the external legs of an amplitude multiplying a much simpler set of scalar-field Feynman diagrams. It is to the latter that the unitarity-based methods may be applied. Earlier applications \cite{Bern:1994zx,Bern:1998ug} of the cutting-rule unitarity methods based on iteration of two-particle cuts gave an expectation that one might have cancellations for $D<10/L + 2$, where $D$ is the spacetime dimension and $L$ is the number of Feynman diagram loops (for $L>1$). Already, this gave an expectation that $D=4$ maximal supergravity might have cancellations of the UV divergences at the $L=3$ and $L=4$ loop orders. This would leave the next significant $D=4$ test at $L=5$ loops. In the ordinary Feynman-diagram approach, a full calculation at this level would involve something like $10^{30}$ terms. Even using the unitarity-based methods, such a calculation would be a daunting, but perhaps not impossible, task.

\begin{figure}[!ht]
\begin{center}
\includegraphics[scale=.5]{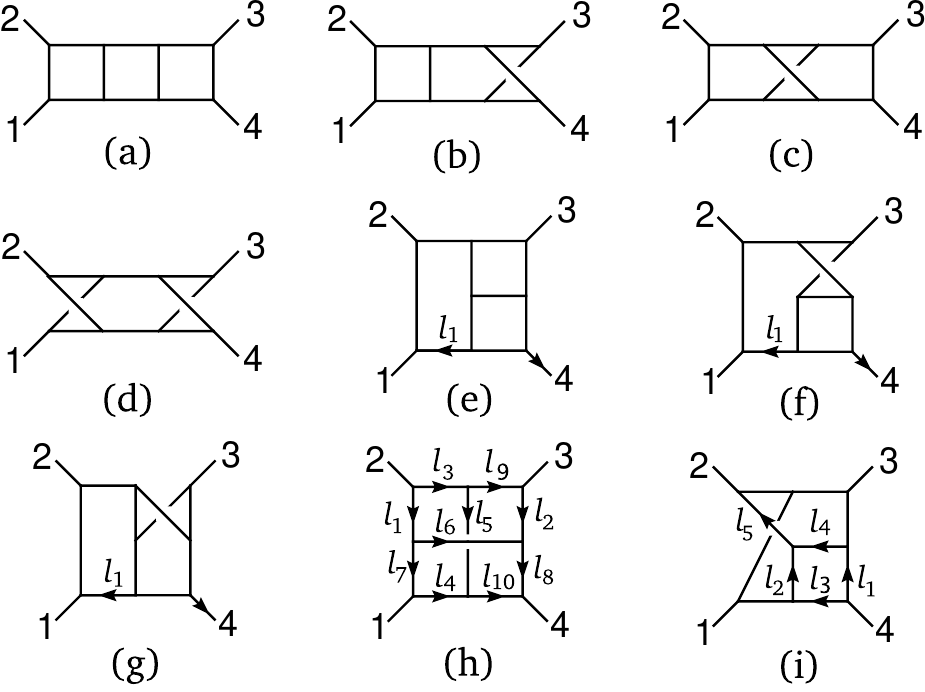} \caption{3-loop Feynman diagram types leading to ultraviolet finiteness of maximal supergravity at this loop order in $D=4$, showing the absence of the $\Delta=8$ $R^4$ divergence. Diagrams (a)-(g) can be analyzed using iterated 2-particle cuts, leading to an expectation of ultraviolet divergence cancellation. Diagrams (h) and (i) cannot be treated this way, but the result nonetheless is a cancellation of the $R^4$ divergence in $D=4$.}\label{fig1}
\vskip.5cm
\includegraphics[scale=.5]{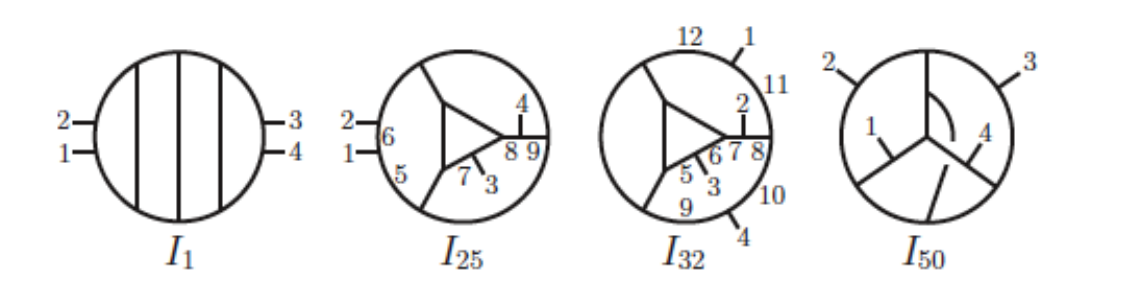} \caption{A sample of 4-loop Feynman diagram types leading to the unexpected ultraviolet finiteness of maximal supergravity at this order in $D=5$. In total, there are 50 such diagram types, with the divergences remarkably summing to zero. This calculation is more revealing in $D=5$ than in $D=4$, because it tests for an apparently allowed $\Delta=14$ $\partial^6 R^4$ divergence.}\label{fig2}
\end{center}
\end{figure}

The striking elements in the 3- and 4-loop calculations of Refs \cite{Bern:2006ew,Bern:2007hh,Bern:2009kd,Bern:2009YM} are the completeness of the calculations and the unexpected further patterns of cancellations found. This has been taken to suggest a possibility of UV cancellations at yet higher loop orders. These results showed that the remaining finite amplitudes display additional cancellations, rendering them ``superfinite''. In particular, earlier work had employed iterated 2-particle cuts and did not consider all diagram types. The complete calculations display further cancellations between diagrams that can be analyzed using iterated 2-particle cuts and additional diagrams that cannot be treated in this way. The nine three-loop diagram types are shown in Figure \ref{fig1}. The end result is that the sum of all diagram types is more convergent by two powers of external momentum than might otherwise have been anticipated. This work was subsequently \cite{Bern:2008pv} reformulated in a way that makes the UV properties manifest diagram-by-diagram.

At the four-loop level \cite{Bern:2009kd}, there are 50 diagram topologies, representative types of  which are shown in Figure \ref{fig2}. The cancellations at this order are more remarkable still. Keeping track of the combinatorics required computerization of the calculation. In the end, the 50 structures combine to yield finite results in both $D=4$ and in $D=5$.

Does such a mechanism cascade to higher-order cases, rendering the maximal N=8 theory completely free of ultraviolet divergences? No one knows at present. Such a scenario might pose puzzling questions for the superstring program, where it has been assumed that ordinary supergravity theories need string ultraviolet completions in order to form consistent quantum theories. On the other hand, there are hints \cite{Berkovits:2006vc,Green:2006yu} from superstring theory that precisely such an all-orders divergence cancellation might take place in the $N=8$ theory. On the other hand, it is not clear exactly what one can learn from superstring theory about purely perturbative field-theory divergences. Examples in Kaluza-Klein theories underline the likely noncommutativity between quantization and truncation of infinite sets of massive states. So, even though supergravity may be obtained as a zero-slope limit of superstring theory at the tree level, it is not completely clear what one can learn about quantized supergravity from superstring theory. A more restrained use of string theory has, however, been used to give arguments suggesting that $D=4, N=8$ supergravity should be finite at least up through six loops \cite{Green:2010sp}. 

One thing that seems clear is that ordinary Feynman-diagram techniques coupled with the nonrenormalization theorems of supersymmetry are unlikely to be able to explain finiteness properties of $N=8$ supergravity at arbitrary loop order. The earlier expectations \cite{Deser:1977nt,Kallosh:1980fi,Howe:1981xy,Howe:1983sr} were that the first divergences unremovable by field redefinitions would occur at three loops in all pure $D=4$ supergravities. A key element in this earlier anticipation was the expectation that the maximal amount of supersymmetry that can be linearly realized in Feynman diagram calculations (\ie ``off-shell supersymmetry'') is half the full supersymmetry of the theory, or 16 out of 32 supercharges for the maximal $N=8$ theory. The precise choice of the half set of supersymmetry generators to be linearly realized does not make an important difference to the result. Thus, for instance, light-cone methods \cite{Mandelstam:1982cb,Brink:1982wv,Kallosh:2008mq} employed a non-Lorentz-covariant quantization technique which, however, maintains manifestly the linear automorphism symmetry ($\textrm{SU}(4)$ for $N=4$ super Yang-Mills, $\textrm{SU}(8)$ for $N=8$ supergravity). Although the intermediate steps are quite different, the allowed counterterms are in the end the same -- although translation between formalisms can be quite involved.

The results of \cite{Bern:2007hh} showed definitely that the expectation of half-BPS operators as the first allowed maximal supergravity and super Yang-Mills counterterms is not sufficiently restrictive. Counterterm analysis quickly came up with an update, however. In \cite{Baulieu:2007ew,Bossard:2009sy} it was shown that there exist non-Lorentz-covariant off-shell formulations with $\ft12$ supersymmetry {\em plus one}, {\it i.e.}\ 9 supercharges for maximal super Yang-Mills and 17 supercharges for maximal supergravity. The super Yang-Mills formulation dimensionally reduces down to (8,1) supersymmetry in $D=2$. In $D=2$ maximal supergravity, there is an analogous formulation with (16,1) supersymmetry. Although a full analysis of this formalism in spacetime dimensions $D>2$ has not been completed, the existence of such a formalism would be enough to push out the boundary of the nonrenormalization theorems so that the half-BPS counterterms (\ref{symf4},\ref{sgr4}) are just {\em ruled out} instead of just being allowed. With this enhanced understanding of the off-shell possibilities, the expectations in various dimensions for the first maximal supergravity divergences would be updated as in Table \ref{tab3}.

\begin{table}[ht]
\centering
\begin{tabular}{|l|c|c|c|c|c|c|c|}
\hline
Dimension $D$\phantom{\Big|}&11&10&8&7&6&5&4\\
\hline
Loop order $L$\phantom{\Big|}&2&2&1&2&3&4&5\\
\hline
BPS degree\phantom{\Big|}&0&0&$\ft12$&$\ft14$&$\ft18$&$\ft18$&$\ft14$\\
\hline
General form\phantom{\Big|}&$\partial^{12} R^4$&$\partial^{10} R^4$&$R^4$&$\partial^4 R^4$&$\partial^6 R^4$&$\partial^6 R^4$&$\partial^4 R^4$\\
\hline
\end{tabular}
\caption{Maximal supergravity divergence expectations based on ``$\ft12$ supersymmetry + 1'' nonrenormalization theorems.
\label{tab3}}
\end{table}

These anticipated enhanced nonrenormalization theorems would dispel some of the mystery of the ``miraculous'' cancellations found in \cite{Bern:2007hh,Bern:2009kd,Bern:2017ucb,Bern:2018jmv}, but this cannot be the full story. The 3-loop cancellations in $D=4$ maximal supergravity would thereby be explained as normal, albeit recondite, consequences of supersymmetry. The 4-loop cancellations in $D=4$ maximal supergravity found in \cite{Bern:2009kd} would similarly be explained. But a different sort of problem poses itself at the 4-loop level in $D=5$.  As one can see from the na\"\i ve degree of divergence given in Eq.\ \eqref{degdiv}, $D=6$, $L=3$ and $D=5$, $L=4$ yield the same na\"\i ve degree of divergence: $\Delta=14$. In dimensional regularization, one only sees divergent expressions corresponding to logarithmic divergences in a standard momentum cutoff regularization. The na\"\i ve degree of divergence thus corresponds to the power counting weight of the fields on the external lines plus the number of derivatives/momenta on those lines. A $\Delta=14$ counterterm for a 4-point operator (the first on-shell nonvanishing structure) corresponds generically to $\partial^6 R^4$ in {\em both} $D=6$, $L=3$ and $D=5$, $L=4$. The $D=6$, $L=3$ divergence is known to occur \cite{Bern:2007hh}, while the $D=5$, $L=4$ structure is now known {\em not} to occur \cite{Bern:2009kd}. The required $D=5$ counterterm might be expected to be formed simply by the dimensional reduction of the corresponding $D=6$ structure. How can such apparently similar structures differ in eligibility under a nonrenormalization theorem?

This question remains a conundrum. But we will see in the next section that additional symmetries of maximal supergravity can yield further constraints on counterterms beyond those arising from the supersymmetry nonrenormalization theorems.

\section{Dualities and renormalization}\label{dualities}

We will concentrate on the impact of duality symmetries in $D=4$ cases, which, although not the most accessible calculationally, are those of the greatest physical interest. The candidate linearized counterterms in $D=4$ maximal supergravity at the three, five and six loop orders are all described by subsurface integrals in the on-shell superspace formalism for the classical equations of motion \cite{Howe:1981xy,Drummond:2003ex,Drummond:2010fp}. These $L=3,\;5,\;6$ invariants all begin with 4-point interactions, of generic structures $R^4$, $\partial^4 R^4$ and $\partial^6 R^4$ and are of BPS degrees 1/2, 1/4 and 1/8 respectively.  They fully account for the 32-supercharge supersymmetry and the linearly realized rigid $SU(8)$ R-symmetry of the $D=4$ theory. They do not, however, fully account for the nonlinear structure of the supersymmetry invariants, nor do they account for the $E_{7(7)}$ duality symmetry which characterizes the classical $N=8$ supergravity theory. As we have seen above, the absence of the $\Delta=8$, $L=3$, $R^4$ divergence can be seen from purely supersymmetry-based field-theoretic arguments \cite{Howe:2002ui,Bossard:2009sy}, results which generalize those for the finiteness of one-half BPS counterterms in maximal super Yang-Mills theories  \cite{Bossard:2009sy}. But the cancellation of the $\Delta=12$, $\partial^4 R^4$ and the $\Delta=14$, $\partial^6 R^4$ divergences is more mysterious.

Now focus attention on the $E_{7(7)}$ duality symmetries. The equations of motion of N=8 supergravity are known to transform under the $E_{7(7)}$ duality symmetry \cite{Cremmer:1979up}, but the action itself is not invariant \cite{Gaillard:1981rj,Aschieri:2008ns}. This problem can be resolved at the cost of manifest Lorenz invariance \cite{Hillmann:2009zf}. It has been shown  \cite{Bossard:2010dq} that $E_{7(7)}$ can be maintained in perturbation theory in $D=4$, although neglecting a possible conflict between supersymmetry and duality symmetry. In particular, duality symmetry becomes anomalous when there is a logarithmic divergence associated to a supersymmetric counterterm that can only be written as a subsurface integral.\footnote{These subtleties were clarified by studying non-duality-invariant supersymmetric counterterms in comparison with the string theory low-energy effective action \cite{Bossard:2014lra,Bossard:2014aea,Bossard:2015uga}. An exhaustive classification of consistent anomalies is nevertheless still lacking, and one resorts to explicit results in string theory and supergravity to fill in the gaps in the proof.} Nonetheless, in loop orders before encountering the first logarithmic divergence, one can safely assume that the candidate counterterm must satisfy both supersymmetry and duality symmetry.  

The $R^4$ invariant (\ref{sgr4}) was shown in Ref.\ \cite{Howe:1981xy} to be invariant at leading order under rigid shift symmetries of the 70 scalar fields, which represent the leading terms in the $E_{7(7)}$ transformations. But it was not clear whether this leading order invariance extends to the full nonlinear theory.  The non-linear supersymmetric invariants have not been constructed explicitly, so in order to check their duality symmetry one must resort to indirect computations.  Using the fact that the tree-level type II string theory effective action includes an $R^4$ type counterterm at order $\alpha'^3$, the five-point and six-point amplitude form-factors with an insertion of $R^4$ were computed in \cite{Brodel:2009hu}. Taking the soft limit for the scalar fields, one can deduce that the tree-level $R^4$ counterterm is not duality invariant, in agreement with the effective Lagrangian. However, this counterterm is not even $SU(8)$ invariant, and as such had no chance to be $E_{7(7)}$ invariant. An $SU(8)$ invariant amplitude was derived in \cite{Elvang:2010kc} by averaging over the different $SU(4)\! \times\!  SU(4)$ scalars  in order to obtain an $SU(8)$ singlet, finding also that the associated counterterm indeed was not  $E_{7(7)}$ duality invariant and is consequently ruled out as a possible divergence structure. This computation was generalized to $\partial^4 R^4$ and  $\partial^6 R^4$ in \cite{Beisert:2010jx}, at the same time as it was verified by direct analysis of the counterterms in \cite{Bossard:2010bd}. The latter argument is relatively simple. One starts from $D>4$ supergravity, where one knows that a specific $f_k(\phi) \partial^{2k} R^4 $ supersymmetry invariant exists, where $f_k(\phi)$ are functions of the scalar fields. For this one may rely on superstring theory using for example Ref.\ \cite{Green:2010wi}, or on the fact that there are respectively 1-loop, 2-loop and 3-loop logarithmic divergences in eight, seven and six dimensions, implying the existence of duality-invariant supersymmetric counterterms of type $R^4$, $\partial^4 R^4$ and $\partial^6 R^4$ \cite{Marcus:1984ei,Bern:2007hh,Bern:2008pv}.\footnote{The first logarithmic divergence in a quantum field theory is associated to a local counterterm satisfying all the Ward identities of the classical action. As such, there must be a duality-invariant and R-symmetry-invariant $R^4$ type counterterm in $D=8$, $\partial^4 R^4$ type in $D=7$ and $\partial^6 R^4$ type in $D=6$. This is indeed consistent with the  low energy effective action in type II string theory \cite{Green:2010wi}.} After dimensional reduction on a torus, one finds therefore a supersymmetry invariant involving a function of the torus volume dilaton in four dimensions. Averaging over $SU(8)$,  one obtains an $SU(8)$ invariant function that can be checked to be non-constant and therefore not duality invariant \cite{Bossard:2010bd}. More generally, it is known that the functions $f_k(\phi)$ satisfy very constraining differential equations \cite{Green:1998by,Green:2010wi,Green:2010kv,Bossard:2014lra,Bossard:2014aea,Bossard:2015uga,Pioline:2015yea}, from which one deduces that $f_k(\phi)=1$ is only possible for $k=0$ in eight dimensions, for $k=2$ in seven dimensions and for $k=3$ in six dimensions, consistently with the supergravity divergences.  To conclude that there is no candidate counterterm satisfying all the necessary symmetries at loop orders $L\le6$, one also needs to prove that there is no other available  supersymmetry invariant. Any non-linear supersymmetry invariant gives an invariant under the linear rigid supersymmetry in the linearised approximation. In four dimensions it is possible to classify the linearised invariants using $SU(2,2|8)$ superconformal  representations, and one concludes that there is a unique quartic invariant for each $R^4$, $\partial^4 R^4$ or  $\partial^6 R^4$  type \cite{Drummond:2003ex}. This linearised analysis is not sufficient by itself to prove that the non-linear invariants exist, but one can rely on superstring theory and logarithmic divergences in supergravity to establish their existence.\footnote{Starting from an existing invariant and acting with $E_{7(7)}$ one obtains the entire set of linearised invariant predicted by the linearised analysis \cite{Bossard:2015uga}, one concludes therefore that there is no obstruction in promoting a linearised invariant to a non-linear invariant. The non-linear invariants are in one-to-one correspondence with the linearised ones.} 

Note that the proof of Ref.\ \cite{Bossard:2010dq} generalises straightforwardly to higher dimensions, provided that there are no Lorentz $\times$ $R$-symmetry 1-loop anomalies.  The absence of such an anomaly is trivial in odd dimensions, and there is none in six dimensions \cite{Marcus:1985yy}. The $SL(2,\bbR)$ symmetry is anomalous at one-loop in eight dimensions, but the latter does not affect the consequences of the tree-level Ward identities for 1-loop divergences, and the $D=8$ $R^4$ counterterm must therefore be associated to an $SL(2,\bbR)$ duality-invariant counterterm in $D=8$.

To summarise the result of this analysis,
including the constraints of duality invariance, and taking into account the presently-known calculational results for loop orders $L\le5$ \cite{Bern:2007hh,Bern:2009kd,Bern:2017ucb,Bern:2018jmv}, the currently anticipated loop orders for supergravity divergence onsets are as follows:
\begin{table}[H]
\centering
\begin{tabular}{|l|c|c|c|c|c|c|c|c|}
\hline
Dimension $D$&11&10&9&8&7&6&5&4\\
\hline
Loop order $L$&{\bf2}&{\bf2}&{\bf2}&{\bf1}&{\bf2}&{\bf3}&6&7\\
\hline
General form&$\partial^{12} R^4$&$\partial^{10} R^4$&$\partial^{8} R^4$&$R^4$&$\partial^4 R^4$&$\partial^6 R^4$&$\partial^{12} R^4$&$\partial^8 R^4$\\
\hline
\end{tabular}
\caption{Current status of known (in boldface) and anticipated maximal supergravity divergence onsets from duality invariance.
\label{tab4}}
\end{table}
\noindent which have been verified to be sharp, \ie to correspond to actual logarithmic divergences, in $D\ge 6$. We certainly expect a logarithmic divergence at 6 loops in five dimensions, but the situation is less obvious in four dimensions. A clearly non-vanishing manifestly $E_{7(7)}$-invariant counterterm certainly becomes available at $L=8$ loops \cite{Howe:1980th,Kallosh:1980fi}. From this analysis, we see that, although the combined constraints of supersymmetry and duality give powerful restrictions on the ultraviolet divergences of maximal supergravity, there is no evidence of cancellation ``miracles'' which might encourage the hope that the maximal theory could remain ultraviolet finite to all loop orders.

Despite this pessimistic view, it was observed that the UV behaviour of maximal supergravity was not worse than the one of maximal super Yang-Mills up to four loops. Up to four loops one indeed observes that the counterterms supporting logarithmic divergences have a power counting $\Delta = L(D-2)+2\ge 8 + 2 L$ for $L\ge 2$, suggesting that the first possible counterterm may need to be of type $\partial^{2L} R^4$. If true, this would imply that the first logarithmic divergence might appear at $L = \frac{6}{D-4}$ loops in all $D<8$ dimensions, suggesting that $N=8$ supergravity might be UV finite  \cite{Bern:2011qn}. Note that this seems consistent with the divergences listed in Table \ref{tab4} in dimension $D\ge 5$. However, the requirement that the counterterm  be of type $\partial^{2L} R^4$ has never been established through a non-renormalisation theorem. This also would have predicted that the five-loop amplitude in dimensional regularisation diverges in $4-\epsilon = \frac{26}{5}$ dimensions, but it has actually been computed to diverge in $4-\epsilon = \frac{24}{5}$ dimensions in \cite{Bern:2018jmv}. One concludes from this that the counterterm of type $\partial^8 R^4$ can formally support a logarithmic divergence at five loops in $D = \frac{24}{5}$ dimensions and the UV behaviour of $N=8$ supergravity turns out to actually be worse that of $N=4$ super Yang-Mills.

In superspace language, the first manifestly $E_{7(7)}$-invariant candidate is simply the volume of superspace, $\int d^4x d^{32}\theta\, \textrm{det} E$  \cite{Howe:1980th}. However, one computes that it vanishes subject to the classical field equation \cite{Bossard:2011tq}. To do this computation, one introduces a harmonic superspace based on $SU(8) / S( U(1)\times U(6)\times U(1))$ and derives a 1/8 BPS harmonic measure for the Grassmann analytic superfields that are annihilated by 4 of the 32 supersymmetries. As an outcome, it follows that the superspace volume vanishes, but one can nonetheless define a different supersymmetric and $E_{7(7)}$ duality invariant counterterm using the 1/8 BPS harmonic measure \cite{Bossard:2011tq}. 

\section{Non-maximal supergravity}
The results of \cite{Bossard:2011tq} extend nicely to all pure supergravity theories with $N\ge 4$ supersymmetry in four dimensions. It appears that the superspace volume vanishes, but one can write a duality-invariant $1/N$ BPS harmonic superspace integral leading to a candidate $(N{-}1)$-loop counterterm of type $\partial^{2(N-4)} R^4$. Using similar techniques, one shows that for loop orders less than $N{-}1$, there is no duality invariant counterterm. The $(N{-}1)$-loop counterterm is therefore the first one satisfying all the necessary symmetries, but it is not a full superspace integral at the nonlinear level.
This nice pattern suggests that the fate of the $(N{-}1)$-loop divergence is the same in all $N$-extended supergravity theories. 

The computation of the three-loop amplitude in $N=4$ supergravity \cite{Bern:2012cd} showed that there is no logarithmic divergence at that order despite the existence of a local counterterm satisfying all the symmetries of the theory. The four-loop amplitude in $N=5$ also turned out to be finite \cite{Bern:2014sna}. So, by extrapolation, one may expect that the $N$-extended (pure) supergravity amplitude is finite at $(N{-}1)$ loops, so that the candidate counterterm described above would never support a logarithmic divergence. 

There is no field-theory explanation for this cancellation based on Ward identities. Before explaining this, let us recall that there is an additional complication in $N=4$ supergravity, because the duality symmetry is anomalous at one loop. The rigid $U(1)$ anomaly computed in  \cite{Marcus:1985yy} gives rise to a 1-loop $SL(2,\bbR)$ anomaly \cite{Bossard:2010dq}. The presence of $U(1)$ violating amplitudes at one loop was explicitly checked in \cite{Carrasco:2013ypa}. Nevertheless, it was shown in \cite{Bossard:2012xs,Bossard:2013rza} that the three-loop counterterm must still be invariant under $SL(2,\bbR)$, so the $R^4$ duality invariant is the only candidate counterterm satisfying all the symmetries.  It was proposed in \cite{Bossard:2012xs,Bossard:2013rza} that the cancellation of the associated divergence could be a consequence of the hypothetical existence of a harmonic superspace off-shell formulation of $N=4$ supergravity, but other consequences of that proposal were shown to be wrong in \cite{Bern:2014lha}, so this off-shell formulation does not exist and there is no satisfactory explanation for the absence of divergence.  

A string-theory explanation for the three-loop finiteness of $N=4$ was proposed in \cite{Tourkine:2012ip}, using the non-renormalisation of the $R^4$ term in the heterotic superstring low-energy effective action. This non-renormalisation was shown to occur at two loops in \cite{DHoker:2005vch}, and then argued in \cite{Tourkine:2012ip} to extend to all orders in perturbation theory. This non-renormalisation is not expected to hold beyond the four-loop order \cite{Tseytlin:1995bi,Bossard:2013rza}, but this provides an explanation for the cancellation of the divergence at three loops found in \cite{Bern:2012cd}. Note nonetheless that this is not really a non-renormalization theorem that one can derive from the symmetries of string theory, but rather it is the result of the explicit computation of a string-theory amplitude, which is consistent with the supergravity computation. 

Pure $N=4$ supergravity was found to diverge at four loops in Ref.\ \cite{Bern:2013uka}. Because the corresponding counterterm also includes a non-duality invariant component, it was proposed in \cite{Bern:2017tuc,Bern:2017rjw,Bern:2019isl} that this divergence could be canceled by introducing a non-duality-invariant counterterm at one loop in order to restore the anomalous $U(1)$ R-symmetry. The string-theory non-renormalisation theorem suggests that the insertion of an $R^2$ counterterm does not spoil the finiteness of the three-loop amplitude, in agreement with the argument provided in \cite{Bern:2019isl}. Whether or not the insertion of an appropriate counterterm can make $N=4$ supergravity finite up to four loops remains to be seen, and there is no indication that such good ultra-violet behaviour should be expected to extend to higher loop orders.

\section{Conclusion}

Let us summarize the current knowledge regarding logarithmic divergences in supergravity. From purely field-theoretic arguments, one can show that  counterterms supporting logarithmic divergences must be invariant under the Cremmer-Julia duality symmetry group, implying that the first candidate counterterm satisfying all the necessary symmetries appears at seven loops in $N=8$ supergravity in four dimensions, and more generally at $N{-}1$ loops in $N$-extended supergravity. The direct computations show that the associated divergence is absent for $N=4$ and $N=5$ supergravities, suggesting that for them the first logarithmic divergence will occur at the $N$-loop order, as was verified explicitly for $N=4$. 

In our discussion, we  have referred several times to string theory, although string theory was only used to provide examples of supersymmetric counterterms in the argument. On the other hand, it is useful to compare the supergravity amplitudes and the  superstring theory amplitudes in the low-energy limit. One can deduce the existence of divergences associated to a BPS protected counterterm of type $\partial^{2k} R^4$ for $k\le 6$ using the superstring effective action as was shown in \cite{Green:2010sp}, and one finds perfect agreement with the supergravity computation.\footnote{Up to a mistake in the three-loop coefficient that was corrected in \cite{Pioline:2015yea,Bossard:2015oxa}.} In particular, for $N=4$ supergravity, the three-loop superstring amplitude is shown to be manifestly finite, whereas the cancellation of the various contributions is very non-trivial in field theory \cite{Bern:2012cd}. This, together with the fact that the $(N{-}1)$-loop counterterm is not the full-superspace integral of a duality-invariant integrand,  gives rise to the hope that there may be a yet-to-be-discovered symmetry explanation for this seemingly miraculous cancellation. 

On the contrary, there is no reasonable symmetry-based argument for the cancellation of the logarithmic divergences for $L\ge N$ in $N$-extended supergravity. It is unlikely that there will ever be a direct computation of the $N=8$ supergravity amplitudes at eight loops, but it would be great to be able to determine indirectly the coefficient of the corresponding divergence. 

\section*{Acknowledgments}
We would like to thank P.\ S.\ Howe for past collaboration on this topic. 
The work of KSS was supported in part by the STFC under consolidated grants ST/T000791/1 and ST/X000575/1.

\bibliographystyle{utphys}


\providecommand{\href}[2]{#2}\begingroup\raggedright\endgroup

\end{document}